\documentclass[10pt,conference]{IEEEtran}

\usepackage{amsfonts}
\usepackage{amsmath}
\usepackage{amssymb}
\usepackage{lscape}
\usepackage{epsf}
\usepackage{graphicx}

\newtheorem{theorem}{\indent Theorem}[section]

\newtheorem{proposition}[theorem]{\indent Proposition}

\newtheorem{EXAMPLE}{\indent Example}[section]
\newtheorem{definition}{\indent Definition}[section]

\newcommand{\code}{{\mathcal{C}}}

\newcommand{\cL}{{\mathcal{L}}}

\newcommand{\cH}{{\mathcal{H}}}

\newcommand{\cF}{{\mathcal{F}}}
\newcommand{\cI}{{\mathcal{I}}}
\newcommand{\cJ}{{\mathcal{J}}}

\newcommand{\cQ}{{\mathcal{Q}}}
\newcommand{\cS}{{\mathcal{S}}}
\newcommand{\cT}{{\mathcal{T}}}
\newcommand{\cU}{{\mathcal{U}}}
\newcommand{\cX}{{\mathcal{X}}}

\newcommand{\sG}{{\mathsf{G}}} 
\newcommand{\sV}{{\mathsf{V}}} 
\newcommand{\sU}{{\mathsf{U}}} 
\newcommand{\sE}{{\mathsf{E}}}

\newcommand{\scp}{{\mathsf{c}}} 
\newcommand{\sfn}{{\mathsf{f}}}

\newcommand\rr{{\mathbb R}}

\newcommand\nn{{\mathbb N}}

\newcommand{\blda}{{\mbox{\boldmath $a$}}}
\newcommand{\bldaa}{{\mbox{\scriptsize \boldmath $a$}}}
\newcommand{\bldb}{{\mbox{\boldmath $b$}}}
\newcommand{\bldbb}{{\mbox{\scriptsize \boldmath $b$}}}
\newcommand{\bldc}{{\mbox{\boldmath $c$}}}

\newcommand{\bldchi}{{\mbox{\boldmath $\chi$}}}

\newcommand{\bldf}{{\mbox{\boldmath $f$}}}
\newcommand{\bldff}{{\mbox{\scriptsize \boldmath $f$}}}
\newcommand{\bldh}{{\mbox{\boldmath $h$}}}

\newcommand{\bldk}{{\mbox{\boldmath $k$}}}
\newcommand{\bldkk}{{\mbox{\scriptsize \boldmath $k$}}}
\newcommand{\bldkappa}{{\mbox{\boldmath $\kappa$}}}
\newcommand{\bldsubkappa}{{\mbox{\scriptsize \boldmath $\kappa$}}}

\newcommand{\bldtau}{{\mbox{\boldmath $\tau$}}}

\newcommand{\bldw}{{\mbox{\boldmath $w$}}}
\newcommand{\bldww}{{\mbox{\scriptsize \boldmath $w$}}}

\newcommand{\bldy}{{\mbox{\boldmath $y$}}}
\newcommand{\bldz}{{\mbox{\boldmath $z$}}}
\newcommand{\bldlambda}{{\mbox{\boldmath $\lambda$}}}
\newcommand{\bldsigma}{{\mbox{\boldmath $\sigma$}}}

\newcommand{\rrr}{\mathfrak{R}}%
\newcommand{\rrrm}{\rrr^{-}}

\newcommand{\Prob}{{p}}

\newcommand{\qed}{\hspace*{\fill}%
    \vbox{\hrule\hbox{\vrule\squarebox{.667em}\vrule}\hrule}\smallskip}
    \def\squarebox#1{\hbox to #1{\hfill\vbox to #1{\vfill}}}

\newlength{\Algwidth}

\begin{document}
\title{Polytope Representations for Linear-Programming Decoding of Non-Binary Linear Codes}   

\author{\authorblockN{Vitaly Skachek}
\authorblockA{Claude Shannon Institute\\
University College Dublin\\
Belfield, Dublin 4, Ireland\\
vitaly.skachek@ucd.ie}
\and
\authorblockN{Mark F. Flanagan}
\authorblockA{DEIS\\
University of Bologna\\
47023 Cesena, Italy\\
mark.flanagan@ieee.org}
\and
\authorblockN{Eimear Byrne$^1$}
\authorblockA{School of Mathematical Sciences\\
University College Dublin\\
Belfield, Dublin 4, Ireland\\
ebyrne@ucd.ie}
\and
\authorblockN{Marcus Greferath\footnotemark$^1$}
\authorblockA{School of Mathematical Sciences\\
University College Dublin\\
Belfield, Dublin 4, Ireland\\
marcus.greferath@ucd.ie}
}

\maketitle

\begin{abstract}
In previous work\footnotetext[1]{These authors are also affiliated with the Claude Shannon Institute for Discrete Mathematics, Coding and Cryptography.}, we demonstrated how decoding of a non-binary linear code could be formulated as a linear-programming problem. In this paper, we study different polytopes for use with linear-programming decoding, and show that for many classes of codes these polytopes yield a complexity advantage for decoding. These representations lead to polynomial-time decoders 
for a wide variety of classical non-binary linear codes. 
\end{abstract}

\vspace{-2ex}
\section{Introduction}

In~\cite{Feldman-thesis} and~\cite{Feldman}, the
decoding of \emph{binary} LDPC codes using linear-programming decoding was proposed, and the connections 
between linear-programming decoding and classical belief propagation decoding were established. In~\cite{FSBG}, the approach of~\cite{Feldman} was extended to coded modulation, in particular to codes over rings mapped to non-binary modulation signals. In both cases, the principal advantage of the linear-programming framework is its mathematical tractability~\cite{Feldman,FSBG}.

For the binary coding framework, alternative polytope representations were studied which gave a complexity advantage in certain scenarios~\cite{Feldman-thesis},~\cite{Feldman},~\cite{Chertkov},~\cite{Feldman-Yang}. Analagous to the work of~\cite{Feldman-thesis},~\cite{Feldman},~\cite{Chertkov},~\cite{Feldman-Yang} for binary codes, we define 
two polytope representations alternative to that proposed in~\cite{FSBG} 
which offer a smaller number of variables and constraints for many classes of nonbinary codes. We compare these 
representations with the polytope in~\cite{FSBG}. These representations are also shown to have equal 
error-correcting performance to the polytope in~\cite{FSBG}.  

\vspace{-1ex}
\section{Linear-programming Decoding}

Consider codes over finite quasi-Frobenius rings (this includes codes over finite fields, but may be more general). 
Denote by $\rrr$ such a ring with $q$ elements, by $0$ its additive identity, and let $\rrrm = \rrr \backslash \{ 0 \}$. Let $\code$ be a linear code of length~$n$ over~$\rrr$ with $m \times n$ parity-check matrix $\cH$. 
 
Denote the set of column indices and the set of row indices of $\cH$ by  $\cI = \{1, 2, \cdots, n \}$ 
and $\cJ = \{1, 2, \cdots, m \}$, respectively. 
The notation $\cH_j$ will be used for the $j$-th row of $\cH$. 
Denote by $\mbox{supp}(\bldc)$ the support of 
a vector $\bldc$. 
For each $j\in\cJ$, let $\cI_j = \mbox{supp}(\cH_j)$ and $d_j=|\cI_j|$, and let $d = \max_{j \in \cJ} \{ d_j \}$. 

Given any $\bldc \in \rrr^n$, parity check $j\in\cJ$ is \emph{satisfied} by $\bldc$ if and only if
the following equality holds over $\rrr$:
\begin{equation}
\sum_{i\in\cI_j} c_i \cdot \cH_{j,i} = 0 \; .
\label{eq:parity_check_satisfied}
\end{equation}
For $j\in\cJ$, define the single parity check code $\code_j$ by 
\[
\code_j = \{ ( b_i )_{i\in\cI_j} \; : \; \sum_{i\in\cI_j} b_i \cdot \cH_{j,i} = 0 \}
\]
Note that while the symbols of the codewords in $\code$ are indexed by $\cI$, the symbols of the codewords in $\code_j$ are indexed by $\cI_j$. 
Observe that $\bldc \in \code$ if and only if all parity checks $j\in\cJ$ are satisfied by $\bldc$.

Assume that the codeword $\bar{\bldc} = (\bar{c}_1, \bar{c}_2, \cdots, \bar{c}_n) 
\in \code$ has been transmitted over a $q$-ary input
memoryless channel,
and a corrupted word $\bldy = (y_1, y_2, \cdots, y_n) \in \Sigma^n$ has been received. Here $\Sigma$ denotes the set of channel output symbols. In addition, assume that all codewords are 
transmitted with equal probability. 

For vectors $\bldf \in \mathbb{R}^{(q-1)n}$, the notation
\[
\bldf = ( \bldf_1 \; | \; \bldf_2 \; | \; \cdots \; | \; \bldf_n ) \; , 
\]
will be used, where
\[
\forall i \in \cI , \; \bldf_i = ( f_i^{(\alpha)} )_{\alpha \in \rrrm} \; .
\]
We also define a function $\bldlambda: \Sigma \longrightarrow ({\mathbb R} \cup \{\pm\infty\})^{q-1}$ by
\[
\boldsymbol{\lambda} = ( \lambda^{(\alpha)} )_{\alpha \in \rrrm}  \; , 
\]
where, for each $y \in \Sigma$, $\alpha \in \rrrm$,
\[
\lambda^{(\alpha)}(y) = \log \left( \frac{\Prob ( y | 0 )}{ \Prob ( y| \alpha ) } \right) \; , 
\]
and $p(y|c)$ denotes the channel output probability (density) conditioned on the channel input. 
Extend $\boldsymbol{\lambda}$ to a map on $\Sigma^n$ 
by $\boldsymbol{\lambda}(\boldsymbol{y})= (\boldsymbol{\lambda}(y_1) \;|\; \boldsymbol{\lambda}(y_2) \;|\; 
\ldots \;|\; \boldsymbol{\lambda}(y_n))$.

The LP decoder in~\cite{FSBG} performs the following cost function minimization:
\begin{equation}
(\hat{\bldf}, \hat{\bldw}) = \arg \min_{(\bldff, \bldww) \in \cQ} \boldsymbol{\lambda} (\bldy) \bldf^T \; ,
\label{eq:object-function}
\end{equation}
where the polytope $\cQ$ is a relaxation of the convex hull of all points $\bldf \in \rr^{(q-1)n}$, which correspond 
to codewords; this polytope is defined as the set of $\bldf\in \mathbb{R}^{(q-1)n}$, together with the auxiliary variables
\[
w_{j,\bldbb} \; \mbox{ for } \;  j\in\cJ, \bldb \in \code_j \; ,
\]
which satisfy the following constraints: 
\begin{eqnarray}
\forall j \in \cJ, \; \forall \bldb \in \code_j,  \quad  w_{j,\bldbb} \ge 0 \; ,
\label{eq:equation-polytope-3} 
\end{eqnarray} 
\begin{equation}
\forall j \in \cJ, \quad \sum_{\bldbb \in \code_j} w_{j,\bldbb} = 1 \; ,
\label{eq:equation-polytope-4} 
\end{equation} 
and 
\vspace{-2ex}
\begin{eqnarray}
&  \forall j \in \cJ, \; \forall i \in \cI_j, \; \forall \alpha \in \rrrm, \nonumber \\
& f_i^{(\alpha)} = 
\sum_{\bldbb \in \code_j, \; b_i=\alpha} w_{j,\bldbb} \; .
\label{eq:equation-polytope-5} 
\end{eqnarray} 

The minimization of the objective function~(\ref{eq:object-function}) over $\cQ$ forms the relaxed LP decoding problem.
The number of variables and constraints for this LP are upper-bounded by $n(q-1) + m q^{d-1}$ and $m(q^{d-1} + d(q-1)+1)$ respectively. 

It is shown in~\cite{FSBG} that if $\hat{\bldf}$ is integral, the decoder output corresponds to the maximum-likelihood (ML) codeword. Otherwise, the decoder outputs an `error'. 


\vspace{-1ex}
\section{New LP Description}

The results in this section are a generalization 
of the high-density polytope representation~\cite[Appendix II]{Feldman}. 
Recall that the ring $\rrr$ contains $q-1$ non-zero elements. Correspondingly, for vectors $\bldk\in\nn^{q-1}$, we adopt the notation
\[
\bldk = (k_{\alpha})_{\alpha\in\rrrm}
\]
Now, for any $j\in\cJ$, we define the mapping
\begin{eqnarray*}
\bldkappa_j \; : \; \code_j & \longrightarrow & \nn^{q-1} \; , \\
\bldb & \mapsto & \bldkappa_j(\bldb)
\end{eqnarray*}
defined by 
\[
( \bldkappa_j (\bldb) )_\alpha = \left| \left\{ i \in \cI_j \; : \; b_i \cdot \cH_{j,i} = \alpha \right\} 
\right|
\]
for all $\alpha \in \rrrm$. We may then characterize the image of $\bldkappa_j$, which we denote by $\cT_j$, as 
\[
\cT_j = \left\{ \bldk \in \nn^{q-1} \; : \; \sum_{\alpha \in \rrrm} \alpha \cdot k_\alpha = 0 \mbox{ and } 
\sum_{\alpha \in \rrrm} k_\alpha \le d_j \right\} \; ,   
\]
for each $j\in\cJ$, where, for any $k\in\nn$, $\alpha\in\rrr$, 
\[
\alpha \cdot k = \left\{ \begin{array}{cc}
0 & \textrm{ if } k = 0 \\
\alpha + \cdots + \alpha & \textrm{ if } k > 0 \textrm{ (} k \textrm{ terms in sum)}\end{array}\right. \; . 
\]
The set $\cT_j$ is equal to the set of all possible vectors $\bldkappa_j(\bldb)$ for $\bldb \in \code_j$. 

Note that $\bldkappa_j$ is not a bijection, in general. We say that a local codeword $\bldb \in \code_j$ is $\bldk$-constrained over $\code_j$ if $\bldkappa_j (\bldb) = \bldk$. 

Next, for any index set $\Gamma\subseteq\cI$, we introduce the following definitions. Let $N = \left| \Gamma \right|$. We define the single-parity-check-code, over vectors indexed by $\Gamma$, by
\begin{equation}
\code_\Gamma = \left\{ \blda = (a_i)_{i\in\Gamma} \in \rrr^N \; : \; \sum_{i\in\Gamma} a_i = 0 \right\} \; . 
\end{equation}
Also define a mapping $\bldkappa_\Gamma \; : \; \code_\Gamma \longrightarrow \nn^{q-1}$ by
\[
\left(\bldkappa_\Gamma(\blda)\right)_\alpha = \left| \left\{ i\in\Gamma \; : \; a_i = \alpha \right\} \right| \; ,  
\] and define, for $\bldk\in\cT_j$, 
\[
\code_\Gamma^{(\bldkk)} = \{ \blda \in \code_\Gamma \; : \; \bldkappa_\Gamma(\blda) = \bldk \} \; .
\]

Below, we define a new polytope for decoding. Recall that 
$\bldy = (y_1, y_2, \cdots, y_n) \in \Sigma^n$ stands for the received (corrupted) word. 
In the sequel, we make use of the following variables:
\begin{itemize}
\item
For all $i \in \cI$ and all $\alpha \in \rrrm$, we have a variable $f_i^{(\alpha)}$. This variable is an indicator of the event $y_i= \alpha$. 
\item
For all $j \in \cJ$ and $\bldk \in \cT_j$, we have a variable $\sigma_{j,\bldkk}$. 
Similarly to its counterpart in~\cite{Feldman}, this variable indicates the contribution to parity check $j$ of 
$\bldk$-constrained local codewords over $\code_j$.  
\item
For all $j \in \cJ$, $i\in\cI_j$, $\bldk \in \cT_j$, $\alpha \in \rrrm$, we have 
a variable $z^{(\alpha)}_{i,j,\bldkk}$. 
This variable indicates the portion of $f_i^{(\alpha)}$ assigned to $\bldk$-constrained local codewords over $\code_j$.
\end{itemize}

Motivated by these variable definitions, for all $j \in \cJ$ we impose the following set of constraints: 
\begin{equation}
\forall i \in \cI_j, \forall \alpha \in \rrrm,  
\qquad f_i^{(\alpha)} = \sum_{\bldkk \in \cT_j} z_{i,j,\bldkk}^{(\alpha)} \; . 
\label{eq:LP_1}
\end{equation}
\begin{equation}
\sum_{\bldkk \in \cT_j} \sigma_{j,\bldkk} = 1 \; .
\label{eq:LP_2}
\end{equation}
\begin{multline}
\forall \bldk \in \cT_j, \forall \alpha \in \rrrm,  \\  
\sum_{i \in \cI_j, \; \beta\in\rrrm, \; \beta\cH_{j,i}=\alpha} z_{i,j,\bldkk}^{(\beta)} = k_\alpha \cdot \sigma_{j,\bldkk} \; . 
\label{eq:LP_3}
\end{multline}
\begin{equation}
\forall i \in \cI_j, \forall \bldk \in \cT_j, \forall \alpha \in \rrrm,
\qquad z^{(\alpha)}_{i,j,\bldkk} \ge 0 \; . 
\label{eq:LP_4}
\end{equation}
\begin{multline}
\forall i \in \cI_j, \forall \bldk \in \cT_j,  \\
\sum_{\alpha \in \rrrm} \; \sum_{\beta\in\rrrm, \; \beta\cH_{j,i}=\alpha} z^{(\beta)}_{i,j,\bldkk} \le \sigma_{j,\bldkk} \; . 
\label{eq:LP_5}
\end{multline}
We note that the further constraints
\begin{equation}
\forall i \in \cI, \forall \alpha \in \rrrm,  
\qquad 0 \le f_i^{(\alpha)} \le 1 \; ,
\label{eq:LP_6}
\end{equation}
\begin{equation}
\forall j \in \cJ, \forall \bldk \in \cT_j,  
\qquad 0 \le \sigma_{j,\bldkk} \le 1 \; , 
\label{eq:LP_7} 
\end{equation}
and
\begin{multline}
\forall j \in \cJ, \forall i \in \cI_j, \forall \bldk \in \cT_j, \forall \alpha \in \rrrm, \quad 
z_{i,j,\bldkk}^{(\alpha)} \le \sigma_{j,\bldkk} \; ,
\label{eq:LP_8} 
\end{multline}
follow from constraints~(\ref{eq:LP_1})-(\ref{eq:LP_5}).
We denote by $\cU$ the polytope formed by
constraints~(\ref{eq:LP_1})-(\ref{eq:LP_5}).  

Let $T = \max_{j \in \cJ} |\cT_j|$.  
Then, upper bounds on the number of variables and constraints in this LP are given by $n(q-1) + m(d(q-1)+1) T$ and 
$m(d(q-1)+1) + m((d+1)(q-1)+d) T$, respectively. Since $T \le {d+q-1 \choose d}$, the number of variables
and constraints are $O(m q \cdot d^{q})$, which, for many families of codes, 
is significantly lower than the corresponding complexity for polytope $\cQ$. 

For notational simplicity in proofs in this paper, it is convenient to define a new set of variables as follows:
\begin{multline}
\forall j \in \cJ, \forall i \in \cI_j, \forall \bldk \in \cT_j, \forall \alpha \in \rrrm, \\ 
\tau^{(\alpha)}_{i,j,\bldkk} = \sum_{\beta\in\rrrm, \; \beta\cH_{j,i}=\alpha} z^{(\beta)}_{i,j,\bldkk} \; . 
\label{eq:tau_definition}
\end{multline}
Then constraints~(\ref{eq:LP_3}) and~(\ref{eq:LP_5}) may be rewritten as
\begin{equation}
\forall j \in \cJ, \bldk \in \cT_j, \forall \alpha \in \rrrm,    
\qquad \sum_{i \in \cI_j} \tau_{i,j,\bldkk}^{(\alpha)} = k_\alpha \cdot \sigma_{j,\bldkk} \; ,
\label{eq:LP_3a}
\end{equation}
\begin{equation}
\forall j \in \cJ, \forall i \in \cI_j, \forall \bldk \in \cT_j, \quad
 0 \le \sum_{\alpha \in \rrrm} \tau^{(\alpha)}_{i,j,\bldkk} \le \sigma_{j,\bldkk} \; . 
\label{eq:LP_5a}
\end{equation}
Note that the variables $\bldtau$ do not form part of the LP description, and therefore do not contribute to its complexity. However these variables will provide a convenient notational shorthand for proving results in this paper.

We will prove that optimizing the cost function~(\ref{eq:object-function}) over this new polytope is equivalent to optimizing over $\cQ$. First, we state the following proposition, which will be necessary to prove this result.

\begin{proposition}
Let $M\in\nn$ and $\bldk\in\nn^{q-1}$. Also let $\Gamma\subseteq\cI$. Assume that for each $\alpha\in\rrrm$, we have a set of nonnegative integers $\cX^{(\alpha)} = \{ x^{(\alpha)}_i \; : \; i\in\Gamma \}$ and that together these satisfy the constraints 
\begin{equation}
\sum_{i\in\Gamma} x^{(\alpha)}_i = k_\alpha M 
\label{eq:lemma-14-req-1}
\end{equation}
for all $\alpha \in \rrrm$, 
and 
\begin{equation}
\sum_{\alpha \in \rrrm} x^{(\alpha)}_i \le M
\label{eq:lemma-14-req-2}
\end{equation} 
for all $i\in\Gamma$. 

Then, there exist nonnegative integers 
$\left\{ w_\bldaa \; : \; \blda \in \code_\Gamma^{(\bldkk)} \right\}$
such that
\begin{enumerate}
\item $\,$ 
\vspace{-3ex}
\begin{equation}
\sum_{\bldaa \in \code_\Gamma^{(\bldkk)}} w_\bldaa = M \; .  
\label{eq:lemma14-claim-1}
\end{equation}
\item
For all $\alpha \in \rrrm$, $i\in\Gamma$, 
\begin{equation}
x^{(\alpha)}_i = 
\sum_{\bldaa \in \code_\Gamma^{(\bldkk)}, \; a_i = \alpha} w_\bldaa \; . 
\label{eq:lemma14-claim-2}
\end{equation}
\end{enumerate}
\label{prop:lemma-14}
\end{proposition}

A sketch of the proof of this proposition will follow at the end of this section. We now prove the main result.

\begin{theorem}
The set $\bar{\cU} = \{ \bldf : \exists \; \bldsigma, \bldz \mbox{ s.t. } (\bldf, \bldsigma, \bldz) \in \cU \}$
is equal to the set $\bar{\cQ} = 
\{ \bldf : \exists \; \bldw \mbox{ s.t. } (\bldf, \bldw) \in \cQ \}$. 
Therefore, optimizing the linear cost function~(\ref{eq:object-function}) over $\cU$ is equivalent to optimizing over $\cQ$. 
\end{theorem}

\emph{Proof:} 
\begin{enumerate}
\item
Suppose, $(\bldf, \bldw) \in \cQ$. For all $j \in \cJ, \bldk \in \cT_j$, we define
\[
\sigma_{j,\bldkk} = \sum_{\bldbb \in \code_j, \; \bldsubkappa_j(\bldbb) = \bldkk} w_{j,\bldbb} \; ,  
\]
and for all $j \in \cJ, \; i \in \cI_j, \; \bldk \in \cT_j$, $\alpha \in \rrrm$, we define
\[
z^{(\alpha)}_{i,j,\bldkk} = 
\sum_{\bldbb \in \code_j, \; \bldsubkappa_j(\bldbb) = \bldkk, \; b_i = \alpha} w_{j,\bldbb} \; ,  
\]

It is straightforward to check that constraints~(\ref{eq:LP_4}) and~(\ref{eq:LP_5}) are satisfied by these definitions. 

For every $j \in \cJ, \; i \in \cI_j, \; \alpha \in \rrrm$, 
we have by (\ref{eq:equation-polytope-5})
\begin{multline*}
f^{(\alpha)}_i \; = \; \sum_{\bldbb \in \code_j, \; b_i \; = \; \alpha} w_{j,\bldbb} \\
    \; = \; \sum_{\bldkk \in \cT_j} \quad \sum_{\bldbb \in \code_j, \; \bldsubkappa_j(\bldbb) = \bldkk, \; b_i = \alpha} w_{j,\bldbb} 
    \; = \; \sum_{\bldkk \in \cT_j} z^{(\alpha)}_{i,j,\bldkk} \; , 
\end{multline*}
and thus constraint~(\ref{eq:LP_1}) is satisfied. 

Next, for every $j \in \cJ$, 
we have by~(\ref{eq:equation-polytope-4})
\begin{multline*}
1 \; = \; \sum_{\bldbb \in \code_j} w_{j,\bldbb} 
  \; = \; \sum_{\bldkk \in \cT_j} \quad \sum_{\bldbb \in \code_j, \bldsubkappa_j(\bldbb) = \bldkk} w_{j,\bldbb} \\
  \; = \; \sum_{\bldkk \in \cT_j} \sigma_{j, \bldkk} \; , 
\end{multline*}
and thus constraint~(\ref{eq:LP_2}) is satisfied. 

Finally, for every $j \in \cJ, \; \bldk \in \cT_j, \; \alpha \in \rrrm$, 
\begin{eqnarray*}
&& \hspace{-8ex} \sum_{i \in \cI_j, \; \beta\in\rrrm, \; \beta\cH_{j,i}=\alpha} z_{i,j,\bldkk}^{(\beta)} \;   \\
  & = & \sum_{i \in \cI_j, \; \beta\in\rrrm, \; \beta\cH_{j,i}=\alpha} \quad 
        \sum_{\bldbb \in \code_j, \; \bldsubkappa_j(\bldbb) = \bldkk, \; b_i = \beta} w_{j,\bldbb} \\
  & = & \sum_{\bldbb \in \code_j, \; \bldsubkappa_j(\bldbb) = \bldkk} \quad 
        \sum_{i \in \cI_j, \; b_i \cH_{j,i} = \alpha} w_{j,\bldbb} \\
  & = & \sum_{\bldbb \in \code_j, \; \bldsubkappa_j(\bldbb) = \bldkk} k_\alpha \cdot w_{j, \bldbb}
  \; = \; k_\alpha \cdot \sigma_{j,\bldkk} \; .
\end{eqnarray*}
Thus, constraint~(\ref{eq:LP_3}) is also satisfied. 
This completes the proof of the first part of the theorem.
\item
Now assume $(\bldf, \bldsigma, \bldz)$ is a vertex of the polytope $\cU$, and so all variables are rational, as are the variables $\bldtau$. 
Next, fix some $j \in \cJ, \bldk \in \cT_j$, and consider the sets
\[
\cX_0^{(\alpha)}= \left\{ \frac{\tau^{(\alpha)}_{i,j,\bldkk}}{\sigma_{j,\bldkk}} 
\; : \; i\in\cI_j \right\} \; . 
\] 
for $\alpha\in\rrrm$. By constraint~(\ref{eq:LP_5a}), for each $\alpha\in\rrrm$, all the values in the set $\cX_0^{(\alpha)}$ are rational numbers between 0 and 1. Let $\mu$ be the lowest common denominator of all the numbers in all the sets $\cX_0^{(\alpha)}$, $\alpha\in\rrrm$. Let 
\[
\cX^{(\alpha)}= \left\{ \mu \cdot \frac{\tau^{(\alpha)}_{i,j,\bldkk}}{\sigma_{j,\bldkk}} 
\; : \; i\in\cI_j \right\} \; ,
\] 
for each $\alpha\in\rrrm$. The sets $\cX^{(\alpha)}$ consist of integers between 0 and $\mu$. By constraint~(\ref{eq:LP_3a}), 
we must have that for every $\alpha \in \rrrm$, the sum of the elements in $\cX^{(\alpha)}$ is equal to $k_\alpha \mu$. 
By constraint~(\ref{eq:LP_5a}), we have 
\[
\sum_{\alpha \in \rrrm} \mu \cdot \frac{\tau^{(\alpha)}_{i,j,\bldkk}}{\sigma_{j,\bldkk}} \le \mu \; 
\]
for all $i \in \cI_j$. 

We now apply the result of Proposition~\ref{prop:lemma-14} with $\Gamma = \cI_j$, $M=\mu$ and with the sets $\cX^{(\alpha)}$ defined as above (here $N=d_j$). Set the variables $\{ w_\bldaa \; : \; \blda \in \code_\Gamma^{(\bldkk)} \}$ according to Proposition~\ref{prop:lemma-14}. 

Next, for $\bldk\in\cT_j$, we show how to define the variables 
$\{ w'_\bldbb \; : \; \bldb \in \code_j, \; \bldkappa_j(\bldb) = \bldk \}$. 
Initially, we set $w'_\bldbb = 0$ for all  $\bldb \in \code_j, \; \bldkappa_j(\bldb) = \bldk$. 
Observe that the values $\mu \cdot z^{(\beta)}_{i,j,\bldkk} /\sigma_{j,\bldkk}$ are 
non-negative integers for every $i \in \cI, \; j \in \cJ, \; \bldk \in \cT_j, \; \beta \in \rrrm$.

For every $\blda \in \code_\Gamma^{(\bldkk)}$, 
we define $w_\bldaa$ words $\bldb^{(1)}, \bldb^{(1)}, \cdots, \bldb^{(w_\bldaa)} \in \code_j$. 
Assume some ordering on the elements $\beta \in \rrrm$ satisfying $\beta \cH_{j,i} = a_i$,
namely $\beta_1, \beta_2, \cdots, \beta_{\ell_0}$ for some positive integer $\ell_0$.
For $i \in \cI_j$, $\bldb_{i}^{(\ell)}$ ($\ell =1 ,2, \cdots, w_\bldaa$) is defined as follows: 
$\bldb_{i}^{(\ell)}$ is equal to $\beta_1$ for the first $\mu \cdot z^{(\beta_1)}_{i,j,\bldkk} /\sigma_{j,\bldkk}$ 
words $\bldb^{(1)}, \bldb^{(2)}, \cdots, \bldb^{(w_\bldaa)}$; $\bldb_{i}^{(\ell)}$ is equal to $\beta_2$ for the next 
$\mu \cdot z^{(\beta_2)}_{i,j,\bldkk} /\sigma_{j,\bldkk}$
words, and so on. 
For every $\bldb \in \code_j$ we define 
\[
w'_\bldbb = \left| \left\{ i \in \{ 1, 2, \cdots, w_\bldaa \} \; : \; \bldb^{(i)} = \bldb \right\} \right| \; . 
\] 

Finally, for every $\bldb \in \code_j, \bldkappa_j(\bldb) = \bldk$, we define
\[
w_{j,\bldbb} = \frac{\sigma_{j, \bldkk}}{\mu} \cdot w'_\bldbb \; . 
\]
 
Using Proposition~\ref{prop:lemma-14}, 
\[
\sum_{\bldaa \in \code_\Gamma^{(\bldkk)}, \; a_i = \alpha} w_\bldaa  = 
\mu \cdot \frac{\tau^{(\alpha)}_{i,j,\bldkk}}{\sigma_{j,\bldkk}} 
= \sum_{\beta \; : \; \beta \cH_{j,i} = \alpha} \mu \cdot \frac{z^{(\beta)}_{i,j,\bldkk}}{\sigma_{j,\bldkk}} \; , 
\]
and so all $\bldb^{(1)}, \bldb^{(2)}, \cdots, \bldb^{(w_\bldaa)}$ (for all $\blda \in \code_\Gamma^{(\bldkk)}$) are well-defined. 
It is also straightforward to see that $\bldb^{(\ell)} \in \code_j$ for $\ell  = 1, 2, \cdots, w_\bldaa$. 
Next, we check that the newly-defined $w_{j,\bldbb}$ 
satisfy~(\ref{eq:equation-polytope-3})-(\ref{eq:equation-polytope-5}) for every $j \in \cJ, \; \bldb \in \code_j$. 

It is easy to see that $w_{j,\bldbb} \ge 0$; therefore~(\ref{eq:equation-polytope-3}) holds. 
By Proposition~\ref{prop:lemma-14} we obtain
\[
\sigma_{j,\bldkk} = \sum_{\bldbb \in \code_j, \; \bldsubkappa_j(\bldbb) = \bldkk} w_{j,\bldbb} \; ,  
\]
for all $j \in \cJ, \bldk \in \cT_j$, and
\[
\tau^{(\alpha)}_{i,j,\bldkk} = 
\sum_{\bldbb \in \code_j, \; \bldsubkappa_j(\bldbb) = \bldkk, \; b_i \cH_{j,i} = \alpha} w_{j,\bldbb} \; ,
\]
for all $j \in \cJ, \; i \in \cI_j, \; \bldk \in \cT_j, \; \alpha \in \rrrm$. 
Let $\beta \cH_{j,i} = \alpha$. 
By the definition of $w_{j,\bldbb}$ it follows that  
\begin{multline*}
\sum_{\bldbb \in \code_j, \; \bldsubkappa(\bldbb) = \bldkk, \; b_i = \beta} w_{j, \bldbb} \; \\
\; = \; \frac{z_{i,j,\bldkk}^{(\beta)}}{\tau^{(\alpha)}_{i,j,\bldkk}} \cdot
\sum_{\bldbb \in \code_j, \; \bldsubkappa(\bldbb) = \bldkk, \; b_i \cH_{j,i} = \alpha} w_{j, \bldbb} 
\; = \; z_{i,j,\bldkk}^{(\beta)} \; , 
\end{multline*}
where the first equality is due to the definition of the words $\bldb^{(\ell)}$, $\ell =1 ,2, \cdots, w_\bldaa$. 

By constraint~(\ref{eq:LP_2}) we have, for all $j \in \cJ$,
\begin{eqnarray*}
1 & = & \sum_{\bldkk \in \cT_j} \sigma_{j,\bldkk} \\ 
  & = & \sum_{\bldkk \in \cT_j} \quad \sum_{\bldbb \in \code_j, \; \bldsubkappa_j(\bldbb) = \bldkk} w_{j,\bldbb} 
  \; = \; \sum_{\bldbb \in \code_j} w_{j,\bldbb} \; ,
\end{eqnarray*}
thus satisfying~(\ref{eq:equation-polytope-4}).

Finally, by constraint~(\ref{eq:LP_1}) we obtain, for all $j \in \cJ, i \in \cI_j, \beta \in \rrrm$,
\begin{multline*}
f_i^{(\beta)} \; = \; \sum_{\bldkk \in \cT_j} z_{i,j,\bldkk}^{(\beta)} \\
   \; = \; \sum_{\bldkk \in \cT_j} \quad \sum_{\bldbb \in \code_j, \; \bldsubkappa_j(\bldbb) = \bldkk, \; b_i = \beta} w_{j,\bldbb}   
   \; = \; \sum_{\bldbb \in \code_j, \; b_i = \beta} w_{j,\bldbb} \; ,
\end{multline*}
thus satisfying~(\ref{eq:equation-polytope-5}). 
\end{enumerate}

\vspace{-1ex}
\subsection*{Sketch of the Proof of Proposition~\ref{prop:lemma-14}}

In this proof, we use a network flow approach (see ~\cite{Cormen}
for background material).

The proof will be by induction on $M$. 
We set $w_\bldaa = 0$ for all $\blda \in \code_\Gamma^{(\bldkk)}$. 
We show that there exists a vector $\blda = \left( a_i \right)_{i \in \Gamma} \in \code_\Gamma^{(\bldkk)}$
such that 
\begin{enumerate}
\item[(i)]
For every $i \in \Gamma$ and $\alpha \in \rrrm$, 
\[
a_i = \alpha \quad \Longrightarrow \quad x^{(\alpha)}_i > 0 \; . 
\]
\item[(ii)]
If for some $i \in \Gamma$, $\sum_{\alpha \in \rrrm} x^{(\alpha)}_i = M$, then 
$a_i = \alpha$ for some $\alpha \in \rrrm$. 
\end{enumerate}

Then, we `update' the values of $x^{(\alpha)}_i$'s and $M$ as follows. For 
every $i \in \Gamma$ and $\alpha \in \rrrm$ with 
$a_i = \alpha$ we set $x^{(\alpha)}_i \leftarrow x^{(\alpha)}_i - 1$. 
In addition, we set $M \leftarrow M - 1$. We also set $w_\bldaa \leftarrow w_\bldaa + 1$. 

It is easy to see that the `updated' values of $x^{(\alpha)}_i$'s and $M$ satisfy  
\[
\sum_{i \in \Gamma} x^{(\alpha)}_i = k_\alpha M
\]
 for all $\alpha \in \rrrm$, 
and $\sum_{\alpha \in \rrrm} x^{(\alpha)}_i \le M$ for all $i \in \Gamma$.
Therefore, the inductive step can be applied with respect to these new values. 
The induction ends when the value of $M$ is equal to zero. 

It is straightforward to see that when the induction 
ter\-mi\-nates, (\ref{eq:lemma14-claim-1}) and~(\ref{eq:lemma14-claim-2}) hold with respect 
to the original va\-lues of the $x^{(\alpha)}_i$ and $M$. 

\subsubsection*{Existence of $\blda$ that satisfies (i)}

We construct a flow network  $\sG = (\sV, \sE)$ as follows: 
$\sV = \{ s, t \} \cup \sU_1 \cup \sU_2$, where $\sU_1 = \rrrm$ and $\sU_2 = \Gamma$. 
Also set
\[
\sE = \{ (s, \alpha) \}_{\alpha \in \rrrm} \; \cup 
\; \{ (i, t) \}_{i  \in \Gamma} 
\; \cup \; \{ (\alpha, i) \}_{x^{(\alpha)}_i > 0} \; .  
\]
We define an integral capacity function $\scp : \sE \longrightarrow \nn \cup \{ + \infty \}$ as follows:
\begin{equation}
\scp(e) = \left\{ 
\begin{array}{cl}
k_\alpha & \mbox{ if } e = (s, \alpha), \; \alpha \in \rrrm \\
1 & \mbox{ if } e = (i, t), \; i \in \Gamma \\
+\infty & \mbox{ if } e = (\alpha, i), \; \alpha \in \rrrm, \; i \in \Gamma
\end{array} \right. \; . 
\end{equation}

Next, apply the Ford-Fulkerson algorithm on the network $(\sG(\sE, \sV), \scp)$ to produce a maximal flow $\sfn_{\max}$. Since 
all the values of $\scp(e)$ are integral for all $e \in \sE$, so 
the values of 
$\sfn_{\max}(e)$ must all be integral for every $e \in \sE$ (see~\cite{Cormen}).

It can be shown that the minimum cut in this graph has capacity $\scp_{\min} = \sum_{\alpha \in \rrrm} k_\alpha$. 

The flow $\sfn_{\max}$ in $\sG$ has a value of $\sum_{\alpha \in \rrrm} k_\alpha$. 
Observe that $\sfn_{\max}((\alpha, i)) \in \{ 0, 1 \}$ for all 
$\alpha \in \rrrm$ and $i \in \Gamma$. 
Then, for all $i \in \Gamma$, we define 
\[
a_i = \left\{ \begin{array}{cl}
\alpha & \mbox{ if } \sfn_{\max}((\alpha, i)) = 1 \mbox{ for some } \alpha \in \sU_1 \\ 
0 & \mbox{ otherwise } 
\end{array} \right. \; . 
\]
For this selection of $\blda = (a_1, a_2, \cdots, a_N)$, we have $\blda \in \code_\Gamma^{(\bldkk)}$
and $a_i = \alpha$ only if $x^{(\alpha)}_i > 0$. 

\subsubsection*{Existence of $\blda$ that satisfies (i) and (ii) simultaneously} 

We start with the following definition. 
\begin{definition}
The vertex $i \in \sU_2$ is called a \emph{critical} vertex, if 
$\sum_{\alpha \in \rrrm} x^{(\alpha)}_i = M$. 
\end{definition}
In order to have~(\ref{eq:lemma-14-req-2}) satisfied after the next inductive step, we have 
to decrease the value of $\sum_{\alpha \in \rrrm} x^{(\alpha)}_i$ by (exactly) 1 
for every critical vertex. This is equivalent to having $\sfn_{\max}((i, t)) = 1$. 

We aim to show that there exists a flow $\sfn^*$ of the same value, which has 
$\sfn^*((i, t)) = 1$ for every critical vertex $i$. 
Suppose that 
there is no such flow. Then, consider the maximum flow $\sfn'$, which has 
$\sfn'((i, t)) = 1$ for the \emph{maximal possible number} of the critical vertices $i \in \sU_2$. 
We assume that there is a critical vertex ${i_0} \in \sU_2$, 
which has $\sfn'(({i_0}, t)) = 0$. 
It is possible to show that the flow $\sfn'$ can be modified towards the flow $\sfn''$ of the same value, 
such that for $\sfn''$ the number of critical vertices $i \in \sU_2$ having 
$\sfn''((i, t)) = 1$ is strictly larger than for $\sfn'$. 

It follows that there exists an integral flow $\sfn^*$ in $(\sG(\sV, \sE), \scp)$ of 
value $\sum_{\alpha \in \rrrm} k_\alpha$,
such that for every critical vertex $i \in \sU_2$, $\sfn^*((i,t)) = 1$. 
We define 
\[
a_i = \left\{ \begin{array}{cl}
\alpha & \mbox{ if } \sfn^*((\alpha, i)) = 1 \mbox{ for some } \alpha \in \sU_1 \\ 
0 & \mbox{ otherwise } 
\end{array} \right. \; . 
\]
and $\blda = (a_i)_{i \in \Gamma}$. For this selection of $\blda$, we have $\blda \in \code_\Gamma^{(\bldkk)}$
and the properties (i) and (ii) are satisfied. 
\qed


\vspace{-1ex}
\section{Cascaded Polytope Representation}

In this section we show that the ``cascaded polytope" representation described in~\cite{Chertkov} and~\cite{Feldman-Yang} can be extended to non-binary codes in a straightforward manner. Below, we elaborate on the details.

For $j \in \cJ$, consider the $j$-th row $\cH_j$ of the parity-check matrix 
$\cH$ over $\rrr$, and recall that 
\[
\code_j = \Big\{ (b_i)_{i \in \cI_j} \; : \; \sum_{i \in \cI_j} b_i \cdot \cH_{j,i} = 0 \Big\} \; .
\]
Assume that $\cI_j = \{ i_1, i_2, \cdots, i_{d_j} \}$ and denote $\cL_j = \{ 1, 2, \cdots, d_j-3 \}$.  
We introduce new variables $\bldchi^j = (\chi^j_i)_{i \in \cL_j}$ and denote $\bldchi = ( \bldchi^j )_{j \in \cJ}$.  

We define a new linear code $\code^{(\chi)}_j$ of length $2d_j-3$ by 
$(d_j - 2) \times (2d_j - 3)$ 
parity-check matrix associated with the following set of parity-check equations over $\rrr$:
\begin{enumerate}
\item $\,$
\vspace{-3ex}
\begin{equation}
b_{i_1} \cH_{j,i_1} + b_{i_2} \cH_{j,i_2} + \chi^j_1 = 0 \; . 
\label{eq:kai-1}
\end{equation}
\item
For every $\ell = 1, 2, \cdots, d_j-4$, 
\begin{equation}
- \chi^j_\ell + b_{i_{\ell+2}} \cH_{j,i_{\ell+2}} + \chi^j_{\ell+1} = 0 \; . 
\label{eq:kai-2}
\end{equation}
\item $\,$
\vspace{-4ex}
\begin{equation}
- \chi^j_{d_j-3} + b_{i_{d_j-1}} \cH_{j,i_{d_j-1}} + b_{i_{d_j}} \cH_{j,i_{d_j}} = 0 \; . 
\label{eq:kai-3}
\end{equation}
\end{enumerate}
We also define a linear code $\code^{(\chi)}$ of length $n + \sum_{j \in \cJ} (d_j-3)$ defined by 
$(\sum_{j \in \cJ} (d_j - 2)) \times (n + \sum_{j \in \cJ} (d_j-3))$ 
parity-check matrix $\cF$ associated with all the sets of parity-check equations~(\ref{eq:kai-1})-(\ref{eq:kai-3}) 
(for all $j \in \cJ$).

\begin{theorem}
The vector $(b_i)_{i \in \cI_j} \in \rrr^{d_j}$ is a codeword of $\code_j$ if and only if there exists some vector 
$\bldchi^j \in \rrr^{d_j-3}$ such that $((b_i)_{i \in \cI_j} \; | \;  \bldchi^j) \in \code^{(\chi)}_j$. 
\label{thrm:code-equivalence}
\end{theorem} 

We denote by $\cS$ the polytope corresponding to the LP relaxation 
problem~(\ref{eq:equation-polytope-3})-(\ref{eq:equation-polytope-5}) for the code 
$\code^{(\chi)}$ with the parity-check matrix $\cF$. 
Let $(\bldb, \bldchi)$ be a word in $\code^{(\chi)}$, where $\bldb \in \code$. 
It is natural to represent points in $\cS$ as $((\bldf, \bldh), \bldz)$, 
where $\bldf = (f_i^{(\alpha)})_{i \in \cI, \; \alpha \in \rrrm}$ 
and $\bldh = (h_{j,i}^{(\alpha)})_{j \in \cJ, \; i \in \cL_j, \; \alpha \in \rrrm}$
are vectors of indicators corresponding to the entries $b_i$ $(i \in \cI)$ in $\bldb$ and 
$\chi^j_i$ $(j \in \cJ, \; i \in \cL_j)$ in $\bldchi$, respectively.

\begin{theorem}
The set $\bar{\cS} = \{ \bldf : \exists \; \bldh, \bldz \mbox{ s.t. } ((\bldf, \bldh), \bldz) \in \cS \}$
is equal to the set $\bar{\cQ} = 
\{ \bldf : \exists \; \bldw \mbox{ s.t. } (\bldf, \bldw) \in \cQ \}$, and
therefore, optimizing the linear cost function~(\ref{eq:object-function}) over $\cS$ is equivalent to optimizing 
it over $\cQ$. 
\label{thrm:polytope-S}
\end{theorem}

It follows from Theorem~\ref{thrm:polytope-S} that 
the polytope $\cS$ 
equivalently describes the code $\code$. This description has at most $n + m \cdot (d-3)$ variables 
and $m \cdot (d-2)$ parity-check equations. However, the number of variables participating in
every parity-check equation is at most $3$. Therefore, the total number 
of variables and of equations in the respective LP problem will be bounded from above by 
\vspace{-1ex}
\[
(n + m (d-3))(q-1) + m(d-2) \cdot q^2  
\] 
\vspace{-1ex}
and 
\vspace{-1ex}
\[
m (d-2) (q^2 + 3q -2)  \; . 
\] 

The polytope representation in this section, when used with the LP problem in~\cite{FSBG}, leads 
to a polynomial-time decoder for a wide variety of classical non-binary codes. Its performance
under LP decoding is yet to be studied. 

\vspace{-1ex}
\section*{Acknowledgements} 

The authors wish to thank O. Milenkovic for interesting discussions. This work was supported 
by the Claude Shannon Institute for Discrete Mathematics, Coding and Cryptography
(Science Foundation Ireland Grant 06/MI/006). 
 

\end{document}